\documentclass[twocolumn]{svjour2}

\usepackage{pslatex}
\usepackage{nicefrac}
\usepackage{latexsym,amsfonts,amsmath,amssymb}
\usepackage{wasysym}
\usepackage{mathptmx}
\usepackage{epsfig}

\title{Evolution of Primordial Black Holes in a radiation and
  phantom energy environment}

\author{Daniel C. Guariento \and J. E. Horvath \and J. A. de
  Freitas Pacheco \and P. S. Custodio}

\institute{D. C. Guariento
\at Universidade de S\~ao Paulo, Instituto de F\'\i sica, S\~ao Paulo,
SP, Brazil \\\email{carrasco@fma.if.usp.br}
\and
J. E. Horvath \and P. S. Custodio
\at Universidade de S\~ao Paulo, Instituto de Astronomia, Geof\'\i
sica e Ci\^encias Atmosf\'ericas, S\~ao Paulo SP, Brazil \\
\email{foton@astro.iag.usp.br}
\and
J. A. de Freitas Pacheco
\at Observatoire de la C\^ote d'Azur, BP 4229, F-06304, Nice Cedex 4,
France
}

\journalname{General Relativity and Gravitation}

\begin{document}

\maketitle

\begin{abstract}

In this work we extend previous work on the evolution of a
Primordial Black Hole (PBH) to address the presence of a dark
energy component with a super-negative equation of state as a
background, investigating the competition between the radiation
accretion, the Hawking evaporation and the phantom accretion, the
latter two causing a decrease on black hole mass. It is found that
there is an instant during the matter-dominated era after which
the radiation accretion becomes negligible compared to the phantom
accretion. The Hawking evaporation may become important again
depending on a mass threshold. The evaporation of PBHs is quite
modified at late times by these effects, but only if the
Generalized Second Law of thermodynamics is violated.



\end{abstract}

\maketitle

\section{Introduction}

The now widely accepted accelerated expansion of the Universe in
its recent history is yet to be fully explained. Several
possibilities to reproduce this effect have been advanced, ranging
from ``conservative'' to very unusual ones requiring new physics.
One of the most economical hypotheses that has received a great
deal of attention is the late dominance of a fluid with an
``anomalous'' equation of state, a sort of analogue of the
inflationary proposals but at a lower energy scale, the so-called
\emph{dark energy}. As is customary to write a generical equation
of state in the form $P = \omega \rho$ and in spite that values of
$\omega$ larger than $-1$ are usually considered, some works have
raised the possibility that the dark sector may be characterized
by a fluid with an equation of state with $\omega < -1$, known
throughout the literature as the \emph{phantom energy}.

There are many physical consequences of such phantom component in
a variety of physical species present in the Universe, most
notably the spacelike singularity known as the \emph{Big Rip}
\cite{caldwell,menace}, or even more fabulous possibilities, like
the \emph{Big Trip} \cite{bigtrip,bigtrip-notes}. Some effort has
been made to remove the Big Rip singularity, but it is still
premature to rule out or support definitely any scenario.

We work within a general phantom energy scenario in this paper. It
has already been acknowledged that, being such an exotic physical
species, the phantom energy may also change the accretion regime
of black holes \cite{babichev-2004}. In the present paper we
investigate the influence of phantom energy accretion onto
primordial black holes (hereafter PBHs) together with the
radiation and matter accretion/evaporation formerly addressed.

The PBH interaction with different types of energy in the universe
is the continuous subject of several sudies, as well as their
interaction with cosmological boundary condidiotns
\cite{harada-03-2007}. Several numerical results also work as test
fields for alternate gravitational theories, and the questions
regarding their very formation at extreme cosmological scenarios are
beginning to yield several interesting results \cite{harada-08-2007}.

We shall focus on the new features specifically introduced by the
phantom era \cite{babichev-2007}, and generally refer to the full
evolution of the PBHs across the mass-time plane. Previous attempts to
address this problem have been limited to the consideration of the
black holes plus phantom fluid only, although there is a more subtle
interplay among components when radiation and matter are also
included (as will be shown below). It is also of interest to
revisit the issue of the black hole behavior in the
radiation-dominated and matter-dom\-i\-nat\-ed eras (that is, well
before the phantom component can be important) for a complete
assessment of the fate of PBHs, especially their behavior in the
matter-dominated era where dark matter may fuel their growth.

\section{PBHs evolution in the early Universe}\label{acrecoes}

\subsection{The radiation-only accretion equation}

Our starting point will be the evolution equation for PBHs in the
radiation-dominated era addressed by several authors (see Cust\'odio
and Horvath \cite{custodio-2002} and references therein), which
takes into account the accretion of radiation and the Hawking
evaporation at a semiclassical level. Ignoring the (potentially
relevant) ``grey factors'' in the absorption of radiation, the
resulting differential equation for the black hole mass $M$ reads
quite generally

\begin{equation}\label{acrecao-simples}
\frac{dM}{dt} = -\frac{A(M)}{M^2} + \frac{27\pi
  G^2}{c^5} \rho_{\mathrm{rad}}(T) M^2
\end{equation}

\noindent with $t$ being the cosmological time, $A(M) = \frac{\hbar
c^4}{G^2} \alpha(M)$, with $\alpha(M)$ called the \emph{running
constant} \cite{cline}, counting the degrees of freedom of the
emitted particles on the Hawking radiation (in CGS units, $A = 7,8
\times 10^{26}$~g$^3$/s for black holes evaporating today
\cite{green}), and $\rho_{\mathrm{rad}}(T)$ the radiation energy
density at temperature $T$ at the time $t$.

In a Universe also filled with phantom energy, the accretion of
such exotic component should also be taken into account. Babichev,
Dokuchaev and Eroshenko \cite{babichev-2004} have worked out a
differential equation for a black hole accreting phantom energy
only, obtaining a counterintuitive result that phantom energy
accretion {\it decreases} the overall black hole mass. The
expression is similar to the accretion term in
eq.~\eqref{acrecao-simples}, and is given by

\begin{equation}\label{acrecao-phantom}
\frac{dM}{dt} = \frac{16\pi G^2}{c^5}
M^2[\rho_{\mathrm{ph}}+p(\rho_{\mathrm{ph}})]
\end{equation}

\subsection{The complete accretion equation}

Considering the radiation accretion and evaporation terms from
eq.~\eqref{acrecao-simples} together with the new phantom energy
accretion term in eq.~\eqref{acrecao-phantom}, and assuming no
interaction between the two different species, the complete
equation for the accretion of the different types of energy into
the black hole is just

\begin{equation}
\frac{dM}{dt} = -\frac{A(M)}{M^2} + \frac{G^2}{c^5}\left[27\pi
  \rho_{\mathrm{rad}}(T) + 16\pi\left(\rho_{\mathrm{ph}} +
  p(\rho_{\mathrm{ph}})\right)\right] M^2
\end{equation}

Using for the phantom energy $p(\rho) = \omega\rho$, $\omega <
-1$, the phantom component of the accretion may be written as

\begin{equation}\label{pressao}
\rho_{\mathrm{ph}} + p(\rho_{\mathrm{ph}}) = (1 + \omega)\rho_{\mathrm{ph}}
\end{equation}

and the complete accretion equation becomes

\begin{equation}\label{acrecao}
\frac{dM}{dt} = -\frac{A(M)}{M^2} + \frac{G^2}{c^5}\left[27\pi
  \rho_{\mathrm{rad}}(T) + 16\pi(1 + \omega) \rho_{\mathrm{ph}}\right]
  M^2
\end{equation}

\subsection{Accretion regimes}

\begin{figure}[!htp]
\centering
\epsfig{file=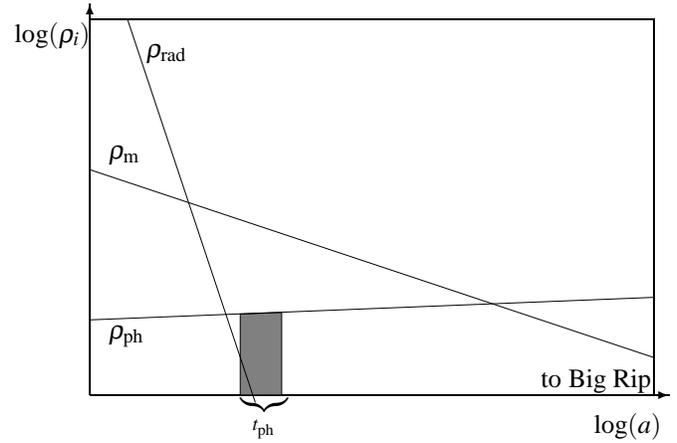}
\caption{Evolution of the radiation, matter and phantom energy
  densities with the scale factor. The exact position of
  $t_{\mathrm{ph}}$ depends on the densities and equation of
  state of the radiation and phantom energy.}\label{evolucao}
\end{figure}

As is well-known, the Friedmann equation can be solved to follow
the cosmological evolution of the phantom energy, as given by
Babichev, Dokuchaev  and Eroshenko \cite{babichev-2004}.

\begin{equation}\label{fried-ph}
\left|\rho + p\right| \propto a^{-3(1+\omega)}
\end{equation}

neglecting all other contributions. The densities of the
radiation, matter and phantom energy terms evolve as represented
in the graphic shown in Figure \ref{evolucao}.

As expected, there is an epoch in which the radiation and phantom
energy accretion terms from eq.~\eqref{acrecao} become comparable.
We call such an epoch the \emph{phantom time}, or
$t_{\mathrm{ph}}$. It must be noted that this instant is distinct
from the one when the lines of Figure \ref{evolucao} cross each
other. The phantom time represents the cosmological instant when
the phantom energy accretion term dominates the radiation term,
changing drastically the black hole evolution dynamics. We can
calculate the value of this time as a function of the initial
radiation and phantom energy densities.

The radiation density as a function of the scale factor is given
by the Friedmann equation, $\rho_{\mathrm{rad}} =
\rho_{\mathrm{rad}}^0\left(\frac{a_0}{a}\right)^4$. During the
matter-dominated era, the scale factor as a function of time is
given by

\begin{equation}\label{eradamateria}
\frac{a(t)}{a_0} = \left(\frac{3H_0t}{2}\right)^{\nicefrac{2}{3}}
\end{equation}

Therefore, the radiation density evolves in the matter-dominated
era as

\begin{equation}\label{evol-rad}
\rho_{\mathrm{rad}} = \rho_{\mathrm{rad}}^0
\left(\frac{3H_0t}{2}\right)^{-\frac{8}{3}}
\end{equation}

Similarly, with the phantom energy eq.~\eqref{pressao} and
evolving according to eq.~\eqref{fried-ph}, and with the time
dependence of the scale factor evolving as of
eq.~\eqref{eradamateria}, the phantom energy density as a function
of time is

\begin{equation}\label{evol-ph}
\rho_{\mathrm{ph}} = \frac{\rho_{\mathrm{ph}}^0}{\left|1 +
  \omega\right|} \left(\frac{3H_0t}{2}\right)^{-{2}
  (1+\omega)}
\end{equation}

The epoch when the phantom energy accretion is as important as the
radiation accretion is the instant when, equating both expressions
according to eq.~\eqref{acrecao}

\begin{equation}\label{densratio}
\rho_{\mathrm{rad}} = -\frac{16}{27}(1 + \omega)\rho_{\mathrm{ph}}
\end{equation}

Inserting the time dependences calculated in eq.~\eqref{evol-rad} and
eq.~\eqref{evol-ph}, this equation yields the \emph{phantom time}.

\begin{equation}\label{t_ph}
\frac{t_{\mathrm{ph}}}{1~\mathrm{s}} = \frac{2}{3H_0}
\left(\frac{16}{27} \frac{\rho_{\mathrm{ph}}^0}
     {\rho_{\mathrm{rad}}^0}\right)^{{\frac{8}{3}} -{2}(1+\omega)}
     \frac{1~\mathrm{km}}{1~\mathrm{Mpc}\cdot 1~\mathrm{s}}
\end{equation}

\noindent with $H_0$ expressed in
$\frac{\mathrm{km}}{\mathrm{s}\cdot\mathrm{Mpc}}$ and, as the
initial values $\rho_{\mathrm{ph}}^0$ and $\rho_{\mathrm{rad}}^0$,
calculated at the end of the matter-dominated era.

We can express this transition time in terms of the redshift,
using eq.~\eqref{densratio}, with the initial conditions
$\rho_{\mathrm{rad}}^0 = 8.12 \times
10^{-13}~\frac{\mathrm{erg}}{\mathrm{cm}^3}$ and
$\rho_{\mathrm{ph}}^0 = 1.79 \times
10^{-8}~\frac{\mathrm{erg}}{\mathrm{cm}^3}$ appropriate for the
obtained conditions, finally coming to $z_{\mathrm{ph}} \simeq 3.1$.

It is reasonable to suppose the transition between radiation and
phantom accretion to be instantaneous due to the very steep
radiation/phantom density ratio, which can be easily seen by
rewriting eq.~\eqref{t_ph} for an arbitrary epoch.

\begin{equation}
\frac{\rho_{\mathrm{rad}}}{\rho_{\mathrm{ph}}} =
\frac{\rho_{\mathrm{rad}}^0}{\rho_{\mathrm{ph}}^0} \left|1 +
\omega\right| \left(\frac{3H_0 t}{2}\right)^{-{\frac{8}{3}} + {2}
(1 + \omega)}
\end{equation}

The radiation density quickly becomes negligible compared to the
phantom energy. The higher the $|\omega|$, the quicker the
transition becomes.

\section{Effects of dark matter accretion}

\subsection{General results}

Up to this point we have neglected completely the possible effects
of (cold) dark matter on the PBHs, which is a popular and
reasonable explanation for the structure formation problem. Within
the CDM scenario, right after the decoupling of dark matter its
accretion onto black holes will depend on the black hole
cross-section for point-like particles. Therefore, the time
dependence of the mass would be given by

\begin{equation}
\frac{dM}{dt} = \frac{16\pi G^2}{c^2}
\frac{\rho{\mathrm{m}}}{u_{\mathrm{m}}} M^2
\end{equation}

\noindent where $u_{\mathrm{m}}$ is the dark matter particle
density, computed after the decoupling

\begin{equation}
u_{\mathrm{m}} \simeq \sqrt{\frac{3k_B
    T_{\mathrm{dec}}}{m}}\frac{1+z}{(1+z)_{\mathrm{dec}}}
\end{equation}

Well before the phantom energy becomes important, the PBH mass
equation, including now the dark matter contribution, is just

\begin{equation}\label{acrecao-dm}
\frac{dM}{dt} = -\frac{A}{M^2} + \frac{27\pi G^2}{c^3}
\rho_{\mathrm{rad}} M^2 + \frac{16\pi G^2}{c^2}
\frac{\rho{\mathrm{m}}}{u_{\mathrm{m}}} M^2
\end{equation}

We must remark that we are always referring to a \emph{diffuse} CDM
component, an appropriate assumption prior to any structure
formation.

\subsection{Numerical predictions}

Because we are interested in the fate of a wide range of black hole
masses, we should integrate equation~\eqref{acrecao-dm} numerically
for several initial conditions and cosmological parameters.

To solve this equation, we first rewrite it in explicitly
time-dependent terms

\begin{equation}
\rho_{\mathrm{m}} = \rho_{\mathrm{m}}^0 (1+z)^3 =
\frac{\rho_{\mathrm{dec}}}{(1+z)_{\mathrm{dec}}^3} (1+z)^3 =
\rho_{\mathrm{dec}} \left(\frac{t_{\mathrm{dec}}}{t}
\right)^{\frac{3}{2}}
\end{equation}

Letting $m = 100~\mathrm{GeV}$ and $T_{\mathrm{dec}} \simeq
1~\mathrm{GeV}$, $(1+z)_{\mathrm{dec}} \simeq 4.26\times 10^{12}$
yields

\begin{equation}
u_{\mathrm{m}} = \sqrt{\frac{3k_B T_{\mathrm{dec}}}{m}}
\frac{1+z}{(1+z)_{\mathrm{dec}}} = 0.173c
\left(\frac{t_{\mathrm{dec}}}{t} \right)^{\frac{1}{2}}
\end{equation}

Setting $\rho_{\mathrm{m}}^0 = \frac{3H_0^2}{8\pi
  G}\Omega_{\mathrm{m}} = 2.24 \times
  10^{-30}~\frac{\mathrm{g}}{\mathrm{cm}^3}$ and
  $\rho_{\mathrm{rad}} = \frac{3}{32\pi Gt^2}$,
  equation~\eqref{acrecao-dm} reads

\begin{equation}
\frac{dM}{dt} = -\frac{A}{M^2} + \frac{81}{32}\frac{G M^2}{c^3}
\frac{1}{t^2} + \frac{16\pi (GM)^2}{c^3} \frac{\rho_{\mathrm{m}}^0
  (1+z)_{\mathrm{dec}}^3}{0.173c} \frac{t_{\mathrm{dec}}}{t}
\end{equation}

which can be solved introducing new variables $y = \frac{M}{M_0}$,
$M_0 = \frac{\alpha c^3 t_0}{G} \def \mathrm{initial black hole mass}$
and $x = \log\left(\frac{t}{t_0}\right)$, yielding an equation of the form

\begin{equation}
\frac{dy}{dt} = -a_1 y^{-2} e^x + a_2 y^2 e^{-x} + a_3 y^2
\end{equation}

\noindent
with $a_1 = \frac{AG}{\alpha c^3 M_0^2} = \frac{1.30 \times
  10^{-13}}{\alpha M_0^2}$, $a_2 = \frac{81}{32}\alpha =
2.53125\alpha$ and $a_3 = 1.38 \times 10^{-42}M_0$.

The dark matter accretion should be taken into account for $x \geq
x_{\mathrm{dec}}$. We may also introduce an instant $x_*$ similar
to the phantom time, in which the dark matter and radiation
accretion have the same value. An estimate for $x_{\mathrm{dec}}$
and $x_*$ is given by

\begin{eqnarray}
t_0 = 2.47\times 10^{-38} M_0 &\rightarrow& x_{\mathrm{dec}} =
\log\left(\frac{6.72 \times 10^{30}}{M_0}\right)\\
a_2 y^2 e^{-x} = a_3 y^2 &\rightarrow& x_* =
\log\left(\frac{1.83 \times 10^{41}}{M_0}\right)
\end{eqnarray}

Table \ref{xis} summarizes a few numerical estimates for
$x_{\mathrm{dec}}$ and $x_*$, covering most of the important PBH
masses.

\begin{table}[!htp]
\centering
\caption{Numerical values for $x_{\mathrm{dec}}$ and $x_0$ for some
  initial values for the black hole mass, along with calculations for
  the times of evaporation with ($x_{\mathrm{evap}}^{\mathrm{m}}$) and
  without ($x_{\mathrm{evap}}$) dark matter.}
\label{xis}
\begin{tabular}{c|cccc}
\hline
$M_0$ (g) & $x_{\mathrm{dec}}$ & $x_*$ & $x_{\mathrm{evap}}$ &
$x_{\mathrm{evap}}^{\mathrm{m}}$\\
\hline
10$^8$ & 52.56 & 76.59 & 63.99 & 63.99\\
10$^9$ & 50.26 & 74.29 & 68.59 & 68.59\\
10$^{10}$ & 47.96 & 71.98 & 73.20 & 73.10\\
10$^{11}$ & 45.65 & 69.68 & 77.81 & 77.81\\
10$^{12}$ & 43.35 & 67.38 & 82.41 & 82.41\\
10$^{13}$ & 41.05 & 65.08 & 87.01 & 87.01\\
10$^{14}$ & 38.75 & 62.77 & 91.62 & 91.62\\
10$^{15}$ & 36.44 & 60.47 & 96.23 & 96.22\\
10$^{16}$ & 34.14 & 58.17 & 100.83 & 100.83\\
\hline
\end{tabular}
\end{table}

An inspection of Table \ref{xis} shows that only black holes with
masses greater than 10$^9$~g should be influenced by the dark
matter accretion at early times. However, this effect of the dark
matter term happens to be small, because it is rapidly overcome by
the accretion of radiation. This can be expected on physical
grounds because the geometrical dilution of the dark matter
component ``starves" the PBHs by quickly diminishing the flux of
particles coming into them. Note that this particular evolution
does {\it not} refer to much later epochs where dark matter halos
had formed, possibly then contributing to the growth of PBHs as
seeds for the ultimate supermassive galactic residents.

The numerical results for the evolution through time are depicted
in Figure \ref{thegraph} for the highest initial condition, as an
example. The resulting bump in the mass (Fig.~\ref{thegraph}) has been
exaggerated for the sake of clarity.

\section{Behavior of the critical mass function}

With expression eq.~\eqref{t_ph} for the time, we can calculate
the value of the critical mass $M_c$ in the instant
$t_{\mathrm{ph}}$. From Cust\'odio and Horvath
\cite{custodio-2002}, the expression for the critical mass is

\begin{equation}\label{massacritica}
M_c(t) \sim 10M_{\textrm{Haw}}
\left(\frac{t}{1\mathrm{s}}\right)^{\frac{1}{2}} \mathrm{g}
\end{equation}

During the late phantom energy accretion dominance era, a critical
mass function would be meaningless, since there is no longer a
relevant mass increase mechanism. Thus, the largest value
reachable by the critical mass in a Universe filled only by
radiation and phantom energy is

\begin{equation}
M_c^{\mathrm{max}} \sim 10M_{\mathrm{Haw}} \frac{2}{3H_0}
     \left(\frac{16}{27} \frac{\rho_{\mathrm{ph}}^0}
     {\rho_{\mathrm{rad}}^0}\right)^{{\frac{8}{3}}
     -{2}(1+\omega)}~\mathrm{g}
\end{equation}

After this time, the Hawking evaporation is no longer a relevant
mechanism for black hole mass decrease, until its mass reaches the
transition value discussed in section \ref{transitiontime}.

It is also convenient to calculate the initial mass of the black
hole which disappears at $t_{\mathrm{ph}}$. For that purpose, it
is enough to consider only the Hawking term in
eq.~\eqref{acrecao-simples}, which yields the well-known solution

\begin{equation}\label{tau}
\tau = \frac{1}{3A(M)}M_i^3
\end{equation}

\noindent where $\tau$ is the evaporation timescale. Restoring the
cgs units $\tau$ reads

\begin{equation}
\tau \sim 10^{71}\left(\frac{M_i}{M_{\mathrm{\astrosun}}}\right)^3
\end{equation}

Combining eq.~\eqref{massacritica} and eq.~\eqref{tau}, we find a
third degree equation in $M_c$, whose solution is the critical
mass of the black hole that will evaporate {\it completely} at $t
= t_{\mathrm{ph}}$

\begin{equation}
\frac{M_c^3}{3A(M)} + \frac{M_c^2}{100M_{\mathrm{Haw}}^2} =
t_{\mathrm{ph}}
\end{equation}

\noindent We use the numerical values of $A(M) \leq 7,8 \times
10^{26}~\frac{\mathrm{g}^3}{\mathrm{s}}$ \cite{custodio-2002} and
$M_{\mathrm{Haw}} \equiv 10^{15}~\mathrm{g}$, as well as the
numerical values of $\rho_{\mathrm{ph}}$, $\rho_{\mathrm{rad}}$,
$\omega$ and $H_0$ necessary to compute $t_{\mathrm{ph}}$. The
instant when the critical mass assumes this value is found by
inverting eq.~\eqref{massacritica}.

Since the mass gain due to radiation accretion is not substantial
\cite{custodio-2002}, all
black holes with $M_i \lesssim M_c^{\mathrm{ph}}$, which reach
critical mass at $T_{\mathrm{cross}}\lesssim t_c$, will disappear
before $t_{\mathrm{ph}}$ and will never reach the phantom era.

\section{The competition between phantom accretion and Hawking
  evaporation}\label{transitiontime}

We have emphasized before that, since after $t_{\mathrm{ph}}$
there is no efficient mechanism that could increase the mass of
the black holes, there is no longer a critical mass function.
However, due to the presence of a phantom field, there are now
\emph{two} distinct regimes of mass decrease, whose relative
importance depends on the mass of a given PBH entering the phantom
era.

Taking eq.~\eqref{acrecao} and neglecting the radiation term, we
can describe the evolution of black hole masses during the phantom
era. Let us define a ratio between the two remaining terms,

\begin{equation}
\xi(M) = \frac{\dot{M}_{\mathrm{ph}}}{\dot{M}_{\mathrm{Haw}}} =
\frac{G^2}{c^3}\frac{16\pi(1+\omega)\rho_{\mathrm{ph}}}{A(M)} M^4
\end{equation}

\noindent or, in terms of a {\it transition mass}

\begin{equation}\label{ratio}
\xi(M) = \left(\frac{M}{M_t}\right)^4
\end{equation}

\noindent
with

\begin{equation}\label{transicao}
M_t = \left[\frac{c^3}{16 \pi G^2}
  \frac{A(M)}{(1+\omega)\rho_{\mathrm{ph}}}\right]^{\nicefrac{1}{4}}
\end{equation}

Substituting numerical values for the constants, we obtain an
expression for $M_t$ in terms of the phantom field density

\begin{equation}\label{transicao-num}
M_t \cong 5.5 \times
10^{17}[(1+\omega)\rho_{\mathrm{ph}}]^{-\nicefrac{1}{4}}~\mathrm{g}
\end{equation}

\noindent with $\rho_{\mathrm{ph}}$ given in g/cm$^3$.

Since both regimes are of mass \emph{decrease}, the black hole
mass will diminish mostly due to phantom accretion until it
reaches $M_t$. After this, the predominant effect will be Hawking
evaporation, since eq.~\eqref{ratio} shows that the change in
regimes is sufficiently sudden for us to make this approximation.

To find the time dependence of the transition mass, we must first
know the evolution of the phantom density. According to the
Friedmann equations for the phantom fluid, we finally obtain
\cite{babichev-2004}

\begin{equation}
(\rho_{\mathrm{ph}})^{-\frac{1}{2}} =
  (\rho_{\mathrm{ph}}^0)^{-\frac{1}{2}} + \frac{3(1+\omega)}{2}
  \left(\frac{8\pi G}{3} \right)^{\frac{1}{2}} t
\end{equation}

\noindent with $(\rho_{\mathrm{ph}}^0)^{-\frac{1}{2}}$ being the
initial density of the phantom field. Inserting this result on
equation eq.~\eqref{transicao-num} the time dependence of $M_t$ is
obtained

\begin{equation}
M_t \cong \frac{8.29 \times
  10^{21}}{(1+\omega)^{\frac{1}{4}}}
  \left[(\rho_{\mathrm{ph}}^0)^{-\frac{1}{2}} + \frac{3(1+\omega)}{2}
  \left(\frac{8\pi G}{3}\right) t \right]^{\frac{1}{2}}~\mathrm{g}
\end{equation}

The initial value of the transition mass depends on both the
initial value of the phantom density and on $\omega$. It is worth
remarking that this transition mass is meaningless in the
radiation-accretion regime.

\begin{figure}[!htp]
\centering
\epsfig{file=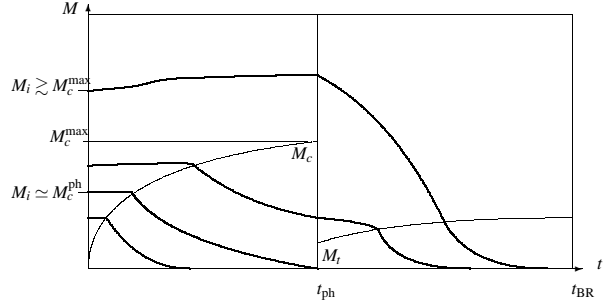,width=.45\textwidth}
\caption{Primordial black hole evolution in the
  matter-radiation-phantom energy scenario. The thick lines represent
  the different trajectories of black holes of different initial
  masses. The Big Rip singularity occurs at $t_{BR}$.}\label{thegraph}
\end{figure}

The differences between the three regimes is depicted in Figure
\ref{thegraph}.

It is important to stress that the Hawking evaporation does not
become negligible after $t_{\mathrm{ph}}$ if taken into account as an
independent process. However, the masses for which it becomes
important ($M < M_t$) drop by a factor of $10^5$ after the
transition. This suddenly drives many black holes, but not all, into
the new regime. When, however, the black holes reach the Planck mass,
a full quantum gravity analysis becomes necessary to properly
determine its fate, since it has been shown that the Hawking
evaporation no longer behaves as expected on such scales
\cite{custodio-2006,coreanos}.

\section{Conclusions}

We have studied in the present work the evolution of PBH for
various regimes of accretion/evaporation in the very early and
contemporary Universe. In particular, we have extended and
clarified the evolution in the radiation-dominated and
matter-dominated eras, including the features of diffuse CDM
accretion producing only a small bump in the mass of the PBHs at
early times. We have generally confirmed previously known features
of the semiclassical pictures of PBH evolution from a general
point of view. Novel features are introduced in this scenario when
a phantom energy component is introduced, as suggested by
Babichev, Dokuchaev and Eroshenko \cite{babichev-2004}.

Broadly speaking, a phantom field introduces another evaporation
regime that competes with the celebrated Hawking evaporation. We
have found that the joint consideration of the relevant terms
quenches the asymptotic approach to a common mass resulting from
the phantom term only. This conclusion should, however, not be
considered as definitive. Its validity rests on the assumption of
the entropy for the phantom fluid being negligible, which is not
the most general possibility. In fact, the enforcement of the
Generalized Second Law (GSL) of thermodynamics would {\it forbid}
the evaporation of the PBHs by phantom accretion
\cite{IzqPav,nosotros} In addition, it is not clear whether the
GSL should be valid in presence of the phantom fluid not
respecting the dominant energy condition, as pointed out by
Izquierdo and Pav\'on \cite{IzqPav}, and models may be constructed
in which the GSL must be modified. There is a rich variety of
behaviors \cite{odintsov,sadjadi} within phantom energy models that remains to
be explored in connection with the PBH evolution problem. In
particular, late evaporation may conflict with the generalized second
law of thermodynamics \cite{nosotros}.

\begin{acknowledgements}

D.C.G and J.E.H. authors wish to thank CNPq (Brazil) for financial
support through grants and fellowships. J.A.F.P has been supported
by CNRS and Fapesp Agency during a research visit to IAG-USP to
complete this work.

\end{acknowledgements}

\end{document}